\documentclass[preprint,12pt]{elsarticle}

\usepackage{graphics}
\usepackage{graphicx}
\usepackage{epsfig}

\usepackage{amssymb}
\usepackage{amsthm}
\usepackage{amsmath}
\usepackage{lscape}
\usepackage{url}

\usepackage{ulem}
\usepackage[dvipsnames]{xcolor} 
\usepackage{float}

\newcounter{bla}

\journal{Computer Physics Communications}

\begin{document}

\begin{frontmatter}



\title{Nexus-CAT: A Computational Framework to Define Long-Range Structural Descriptors in Glassy Materials from Percolation Theory}

\author[a]{Julien Perradin\corref{author}}
\author[a]{Simona Ispas}
\author[a,b]{Anwar Hasmy}
\author[a]{Bernard Hehlen}

\cortext[author] {Corresponding author.\\\textit{E-mail address:} julien.perradin@protonmail.com}
\address[a]{Laboratoire Charles Coulomb, CNRS - University of Montpellier, 34095 Montpellier, France.}
\address[b]{Departamento de F\'{i}sica, Universidad Sim\'on Bol\'{i}var, Valle de Sartenejas, Caracas, Venezuela}

\begin{abstract}




Nexus-CAT (Cluster Analysis Toolkit) is an open-source Python package for cluster detection and percolation analysis of atomistic simulation trajectories. 
Standard structural tools, such as the pair distribution function or structure factor, fail to capture the long-range connectivity changes underlying amorphous-amorphous transitions in glassy materials. 
Nexus-CAT addresses this gap by reading extended XYZ trajectory files and identifying clusters via a Union-Find algorithm with path-compression.
Four clustering strategies, i.e., distance-based, bonding, coordination-filtered, and shared-neighbor, are implemented through a Strategy Factory design pattern, enabling the treatment of diverse network topologies.
The program computes key percolation properties
with percolation detection based on a rigorous period vector algorithm. 
The package is validated against theoretical predictions and applied to glasses with different bonding environments, namely vitreous silica, vitreous ice, and amorphous silicon.
One original result is the observation of a percolation transition prior to crystallization in the latter, indicating that pressure-induced crystallization is initially driven by an amorphous transformation with similar coordination number.
The code is also designed to be readily extended to gels, cements, and other disordered materials.
Nexus-CAT is fully available on GitHub and PyPI.
\\


\noindent \textbf{PROGRAM SUMMARY/NEW VERSION PROGRAM SUMMARY}

\begin{small}
\noindent
{\em Program Title: }   Nexus-CAT (Cluster Analysis Toolkit)                                       \\
{\em CPC Library link to program files:} (to be added by Technical Editor) \\
{\em Developer's repository link:} https://github.com/jperradin/nexus \\
{\em Licensing provisions(please choose one):} MIT  \\
{\em Programming language:} Python                                  \\
{\em Supplementary material:}                                 \\
{\em Journal reference of previous version:}*                  \\
{\em Does the new version supersede the previous version?:}*   \\
{\em Reasons for the new version:*}\\
{\em Summary of revisions:}*\\ \\
{\em Nature of problem(approx. 50-250 words):}\\ 
When analysing the structure of disordered systems, standard structural analysis tools such as the pair distribution function, bond angular distribution, or structure factor provide local or reciprocal-space descriptions of the structure. However, owing to structural disorder, they fail to capture long-range emergent order, e.g., scale-invariant descriptors analogous to the order parameter of crystalline phase transitions.  
Recent studies have proposed percolation theory as a unified framework for describing amorphous-amorphous transformations, showing that the formation and breakdown of coordination-specific polyhedral networks follow universal scaling laws characteristic of percolation phase transitions~\cite{hasmy_percolation_2021,hasmy_unravelling_2024,perradin_criticality_2025,ioannidou_crucial_2016}.
This percolation-based description introduces a genuinely long-range view of system-spanning networks in glasses that extends
beyond the reach of traditional short- and medium-range order analyses.
However, to our knowledge, no flexible and dedicated tools currently exist for performing systematic cluster detection and percolation analysis on sequences of point‑like particle configurations, like atomistic trajectories generated by Monte Carlo or molecular dynamics simulations.
\\ \\
{\em Solution method(approx. 50-250 words):}\\
Nexus-CAT addresses this gap by providing an open-source Python framework that reads standard extended XYZ trajectory files and identifies atomic clusters using a high-performance Union–Find algorithm. 
The toolkit implements multiple clustering strategies, including distance-based, bonding, coordination-filtered, and shared-neighbor approaches, enabling the treatment of diverse disordered network topologies.
This versatility makes Nexus-CAT applicable to a wide range of disordered systems, from network-forming to multicomponent oxide glasses, as well as chalcogenide and metallic glasses, amorphous semiconductors, amorphous ice, etc.
The pipeline provides rigorous computation of percolation properties (\textit{e.g.}, correlation length, order parameter, average cluster size).
Combined with finite-size scaling analyses over multiple system sizes, these quantities allow the characterization of the critical behavior and universality class governing amorphous-amorphous transitions, as recently demonstrated for vitreous silica and amorphous ice~\cite{perradin_criticality_2025,perradin_polyamorphism_2025,hasmy_unravelling_2024}.
This framework can be readily extended to other systems, such as gels and cements~\cite{ioannidou_crucial_2016, nabizadeh_network_2024}, thereby enabling a systematic investigation of polyamorphic transitions responsible for macroscopic changes across a broad class of disordered materials.
%
\end{small}
\end{abstract}
\end{frontmatter}


\section{Introduction}
\label{sec:introduction}

Some amorphous and disordered materials undergo structural transitions under varying thermodynamic conditions, commonly referred to as polyamorphic or amorphous-amorphous transitions (AATs)~\cite{mcmillan_polyamorphic_2004,wilding_structural_2006,machon_pressure-induced_2014,mishima_melting_1984,mishima_polyamorphism_2010,brazhkin_high-pressure_2003}, that remain challenging to characterize at the atomic scale.
In \textit{v}-SiO$_2$, for instance, compression induces a progressive transformation of the tetrahedral SiO$_4$ network into higher-coordinated species (SiO$_5$, SiO$_6$), fundamentally altering the material's structure and mechanical properties~\cite{hasmy_percolation_2021,perradin_polyamorphism_2025}.
Similar phenomena are observed in other disordered systems, including amorphous ice~\cite{hasmy_unravelling_2024} and amorphous silicon~\cite{deringer_origins_2021}.
Understanding the nature and universality of these transitions is a central question in condensed matter physics.

The standard structural tools typically employed to analyze atomistic simulation trajectories such as the pair distribution function, bond angular distribution, and structure factor, probe short- to medium-range order and are therefore not well suited to capture the long-range connectivity changes that govern these transitions~\cite{hasmy_percolation_2021,perradin_polyamorphism_2025,hasmy_unravelling_2024}.
Recent studies have demonstrated that percolation theory provides an appropriate framework to describe AAT, revealing that the formation and breakdown of polyhedral networks follow universal scaling laws characteristic of continuous phase transitions~\cite{perradin_criticality_2025,perradin_polyamorphism_2025,hasmy_unravelling_2024}.
This percolation-based approach introduces a genuinely long-range structural descriptor in glasses that complements and extends beyond traditional local-order analyses.

To the best of our knowledge, no open‑source tool currently provides a general framework for cluster‑based, multi‑species percolation analysis directly from atomistic trajectories of disordered systems.
General-purpose graph libraries, such as NetworkX~\cite{hagberg_exploring_2008}, and analysis packages, such as OVITO~\cite{stukowski_visualization_2009}, provide basic capabilities for cluster identification.
Existing percolation‑specific tools are either tailored to gel‑point detection in cross‑linking polymer networks (\textit{e.g.}, percolation-analyzer)~\cite{livraghi_exact_2021} or to lattice percolation and ionic transport in crystalline solids (\textit{e.g.}, Dribble)~\cite{noauthor_atomisticnetdribble_2025,urban_configurational_2014,lee_unlocking_2014,ouyang_effect_2020}.
However, none of these tools provides the integrated workflow required for percolation studies in off-lattice, multi-species disordered systems, namely, the combination of physically motivated clustering strategies, and the rigorous computation of the full set of scaling percolation properties, \textit{e.g.}, correlation length, order parameter, average cluster size, and percolation probability.

To address this gap, we introduce Nexus-CAT (Cluster Analysis Toolkit), an open-source Python package for the post-processing, identification, and percolation-based analysis of atomistic simulation trajectories.
The toolkit implements multiple clustering strategies tailored to the physics of disordered network-forming materials, and computes a comprehensive set of percolation properties with statistical averaging over trajectory frames.
Developed primarily for atomistic systems, Nexus-CAT provides a flexible and extensible framework that can be readily adapted to a wide range of other complex systems.
The source code is available on GitHub at \url{https://github.com/jperradin/nexus}, on PyPI, and the complete documentation on ReadTheDocs.

\section{Methodology}
\label{sec:methodology}

\begin{landscape}
\begin{figure*}[htbp]
    \centering
    \includegraphics[width=\linewidth]{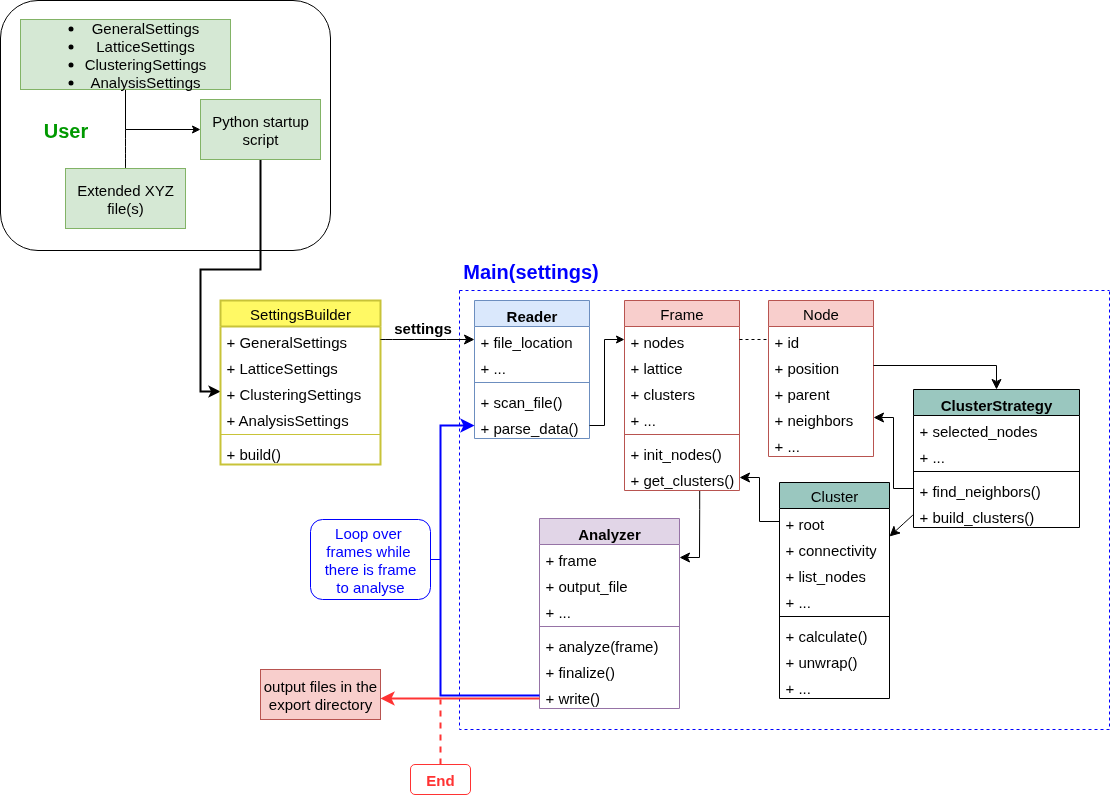}
    \caption{Typical workflow of Nexus-CAT python package (attribute and method names are not those used in the program for the sake of visibility, refer to package documentation if needed).
    }
    \label{fig:workflow}
\end{figure*}
\end{landscape}

In this section, we introduce the \textit{main} function of the program.
To illustrate its general procedure, the typical workflow is depicted in Fig.\ref{fig:workflow}.
The \textit{main} function is executed through a Python launch script that contains all necessary information (\textit{e.g.}, file location, export directory) as well as the clustering and analysis parameters.
The user defines these parameters in four dedicated blocks: \textit{GeneralSettings}, \textit{LatticeSettings}, \textit{ClusteringSettings}, and \textit{AnalysisSettings}.
They are then processed by the \textit{SettingsBuilder} block (highlighted in yellow in Fig.~\ref{fig:workflow}), which constructs a \textit{Singleton} settings object.
This \textit{settings} object is subsequently propagated through each workflow block when necessary.
The structure and role of the settings object will be discussed in detail in Sec.~\ref{sec:settings}, following the introduction of the program’s core components (see the colored blocks in Fig.\ref{fig:workflow}).

The program only requires a trajectory XYZ file containing node types (corresponding to elementary units such as single atoms, atomic arrangements, beads used in coarse-grained simulations, or even trees in fire propagation models, etc.) with their associated XYZ coordinates, and the user-specified settings in the Python launch script.

\subsection{Reader and data importation}

The Reader block (blue block in Fig.~\ref{fig:workflow}), called by the \textit{main} function, uses the relevant configuration options: the input file path, the list of frames to read, and other parameters specified in \textit{GeneralSettings}, together with any lattice information supplied in \textit{LatticeSettings}.
Note that the lattice is defined by a full $3\times 3$ matrix (with components $l_{xx}, l_{xy}, l_{xz}, \dots$), which enables the code to handle general parallelepipeds. 
The frames are handled sequentially in a \textit{for} loop; at each iteration, the \textit{Reader} jumps to the appropriate frame, reads its contents, and generates a \textit{Frame} object (red block in Fig.~\ref{fig:workflow}).

Each \textit{Frame} contains the list of atoms or particles imported from the XYZ trajectory file, and each particle/atom is represented as an instance of the \texttt{Node} class (red block \textit{Node} in Fig.~\ref{fig:workflow}). 
This class stores essential attributes, such as the {atom}'s species and position, which are systematically wrapped in the lattice provided by the trajectory file in the comment line of each frame
or by the user, its parent in the Union–Find data structure (see below), and its list of neighbors. These attributes are fundamental to the high-performance Union–Find algorithm used for cluster detection~\cite{galler_improved_1964} (see Sec.~\ref{sec:union_find}).

\subsection{Neighbors search}

Once a \textit{Frame} object is created, it is associated with all the \textit{Node} objects.
Each \texttt{Node} instance has the \texttt{neighbors} attribute.
This attribute is a list of \textit{Node} objects surrounding the central atom or node.
Then, the nearest neighbors are found by the \textit{ClusterStrategy} block (turquoise block in Fig.~\ref{fig:workflow}) using the K-D tree algorithm via the \texttt{cKDTree} function from the SciPy package \cite{virtanen_scipy_2020}.
A \textit{for} loop traverses each networking atom and finds the neighbors below the user-specified cut-off distance in \textit{ClusteringSettings}.
Note that periodic boundary conditions (PBC) are enabled by default and can be disabled in the \textit{GeneralSettings}. 
If there are more than one species in the system, neighbors are filtered using a set of pairwise cut-off distances provided by the user in the settings. 

\subsection{Cluster identification: The Union-Find algorithm}
\label{sec:union_find}

Once the nodes and their neighbors have been identified, the system's local structure can be modeled as a graph in which each \texttt{Node} object is connected to its neighbors if they satisfy the specified criteria (e.g., a cutoff distance). To identify the connected components (clusters) within this graph effectively implementing the \textit{Friend-of-a-Friend} clustering model~\cite{huchra_groups_1982}, we employ the \textbf{Union-Find} (or Disjoint-Set) data structure~\cite{galler_improved_1964}.

This algorithm is utilized by the \textit{ClusterStrategy} class to efficiently manage and merge sets of atoms. It operates on \texttt{Node} objects using two core operations:

\begin{itemize}
    \item \texttt{find(node)}: Determines the representative (or root) of the set to which a node belongs. Our implementation uses \textit{path compression}, a technique that flattens the tree structure during traversal by linking nodes directly to the root. This results in significant efficiency gains for subsequent operations.
    \item \texttt{union(node1, node2)}: Merges the sets containing two nodes. It first identifies the roots of both nodes; if they differ, it links one root to the other, thereby unifying the two clusters.
\end{itemize}

Based on the chosen strategy (see Section \ref{sec:clustering_strat}), the \texttt{union} method is iteratively called for every pair of nodes satisfying the connectivity criteria.
Once all connections are processed, the program performs a final pass over all nodes using the \texttt{find} method to group them by their unique root parent. Finally, a \texttt{Cluster} object (represented as the turquoise block in Fig.~\ref{fig:workflow}) is instantiated for each group.
Note that isolated nodes are ignored in this analysis; consequently, the program reports a minimum cluster size of 2.

Finally, it is worth noting that other graph traversal algorithms, such as Breadth-First Search (BFS)~\cite{moore_shortest_1959} or Depth-First Search (DFS)~\cite{tarjan_depth-first_1972}, can yield similar clustering results. However, the Union-Find approach was preferred in this work for its algorithmic efficiency and its natural integration within Nexus-CAT's object-oriented architecture.

\subsection{Clustering Strategies}
\label{sec:clustering_strat}

To accommodate the diverse physics of various systems, Nexus-CAT employs a Strategy Factory design pattern.
This factory inspects the user-provided \textit{ClusteringSettings} object and dynamically selects the most appropriate clustering algorithm for the analysis.

\begin{figure}[!htbp]
    \centering
    \includegraphics[width=0.7\linewidth]{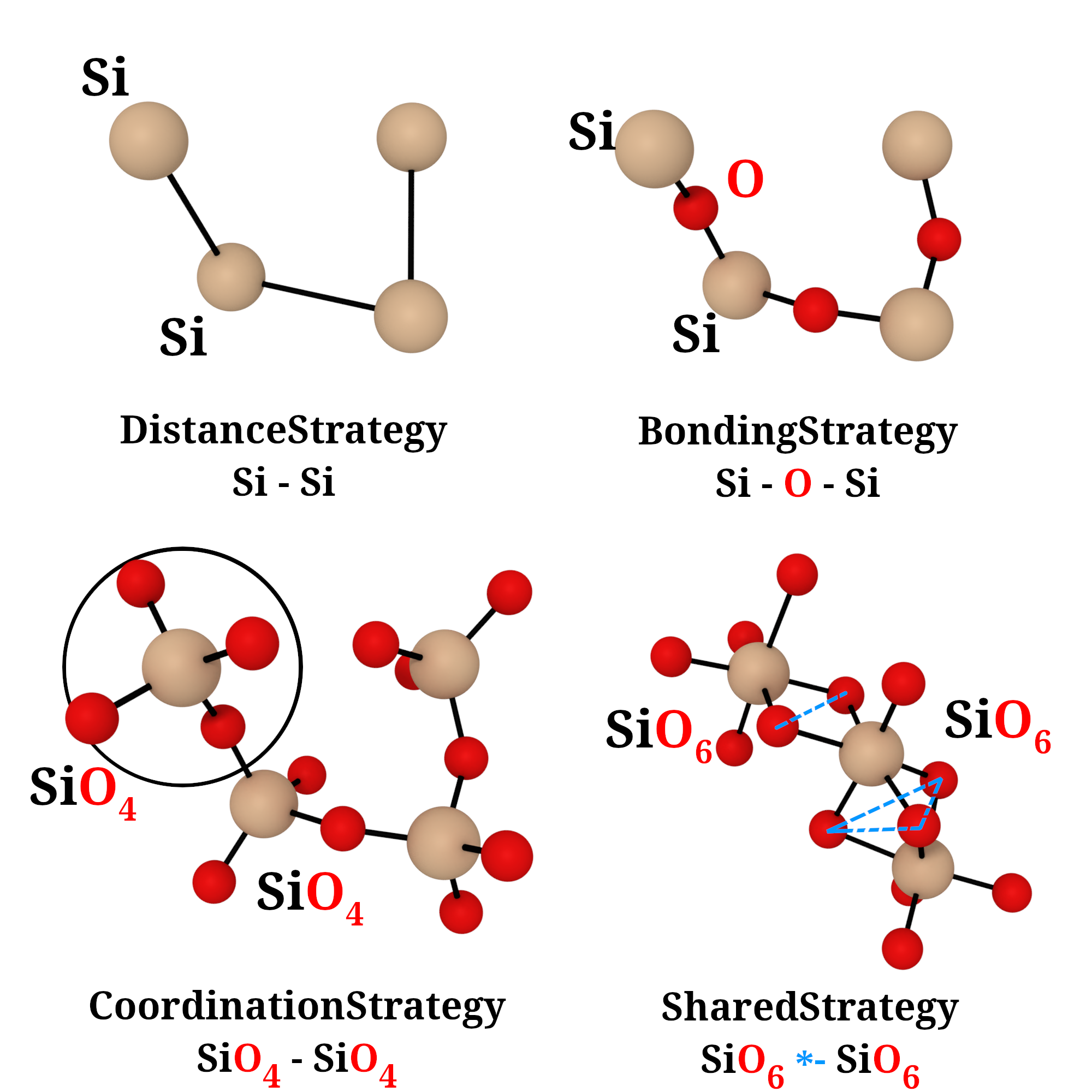}
    \caption{Representation of the different strategies applied to amorphous SiO$_2$. \textit{Top-left} is DistanceStrategy, which simply connects Si-Si based on a single cutoff. \textit{Top-right} is BondingStrategy, which connects two Si atoms bridged by a common O atom. \textit{Bottom-left} is \textit{CoordinationStrategy} which first identifies SiO$_4$ units and then connects those units if they share an oxygen. \textit{Bottom-right} is \textit{SharedStrategy}, which is similar to \textit{CoordinationStrategy} but adds a threshold to the number of shared common neighbors, \textit{i.e.}, SiO$_6$ sharing two or more common Oxygens.}
    \label{fig:strategies}
\end{figure}

In the following, we illustrate the core strategies using amorphous SiO$_2$ as an example, where the primary nodes are Si atoms and the neighbors are O atoms.
However, these quantities are not hardcoded; in practice, Si and O can be replaced by any particle type or atomic species appropriate to the system under study. 
For instance, in amorphous ice, one could define the oxygens of water molecules (H$_2$O) as the primary nodes and analyze their network using the same framework (see Sec.~\ref{sec:non-bonded_aH2O}).
The core strategies, currently available and visualized in Fig.~\ref{fig:strategies}, are as follows:

\begin{itemize}
    \item \textbf{Distance Strategy}: This is the most direct approach, forming clusters by connecting any two nodes that are within a specified cut-off distance.
    \item \textbf{Bonding Strategy}: This strategy identifies networks based on two primary nodes only if they share a common neighbour of a specified species.
    \item \textbf{Coordination Strategy}: Building upon the other strategies, this method adds a layer of specificity by filtering connections based on the coordination number $Z$ of the nodes. This enables a more nuanced analysis of the network structure.
    For instance, in amorphous SiO$_2$, it can be configured to connect only four-coordinated silicon atoms, thus specifically analysing the SiO$_4$ tetrahedral network. This strategy supports several search modes to define the desired connectivity precisely:
    \begin{itemize}
        \item Default mode: Connects nodes if their coordination numbers fall within a specified range (\textit{e.g.}, connecting all Si atoms with $Z$ between 5 and 20 to calculate all other polyhedra than SiO$_4$)
        \item Pairwise mode: Creates clusters of nodes with identical coordination number (e.g., SiO$_4$-SiO$_4$ or SiO$_5$-SiO$_5$).
        \item Alternating Mode: it also forms clusters between nodes with identical coordination numbers as the pairwise mode does, but also with the coordination number of the second node being one greater than the first, such as SiO$_4$-SiO$_5$, SiO$_5$-SiO$_6$...
        \item Mixing Mode: Forms clusters between nodes with all possible pairs of coordination numbers specified in the range.
    \end{itemize}
    \item \textbf{SharedStrategy}: A constraint is added to the \textit{CoordinationStrategy} condition: a connection is only formed if the two nodes also share a minimum number of common neighbors, as defined by the \texttt{shared\_threshold} setting. This enables the identification of complex structural motifs, such as distinguishing among corner-sharing (1 shared neighbor), edge-sharing (2 shared neighbors), and face-sharing (3 or more shared neighbors) polyhedra.
\end{itemize}

Once clusters have been identified, the \textit{ClusterStrategy} block calls the \textit{Analyzer} block (purple block in Fig.~\ref{fig:workflow}), which computes one cluster/percolation property based on cluster connectivity. 
Below is a list of key computed properties and their corresponding technical details.
Several properties can be selected in the \textit{AnalysisSettings} block, and the code then calls the appropriate \textit{Analyzer} block for each property via the \textit{AnalyzerFactory}.

\begin{itemize}
    \item [a)] The concentration $\phi$ represents the proportion of nodes that participate in cluster formation for a given connectivity type, relative to the total number of nodes in the system. It is a fundamental metric and is equivalent to the probability parameter $p$ in percolation theory.
    \item[b)] The average cluster size $\langle S\rangle$ is computed as the second moment of the average. This critical quantity diverges at the percolation threshold, which is equivalent to a susceptibility in thermal phase transition~\cite{stauffer_introduction_2018}. To focus on the finite cluster distribution, the percolating cluster is excluded from this calculation.
\begin{equation}
    \langle S(\phi) \rangle = \frac{\sum_s^* s^2 n_s(\phi)}{\sum_s^* s n_s(\phi)}
\end{equation}
where $s$ refers to the size, \textit{i.e.}, the number of sites belonging to the cluster, $n_s$ refers to the number of clusters of size $s$, and the sum $\sum_s^*$ is exclusive to finite clusters.
\item [c)] The gyration radius $R_{g,s}$ measures a cluster's spatial extent and is calculated as the root mean square distance of the nodes from the cluster's center of mass, $\vec{R}_{\text{com}}$, which is defined as the average position of all node coordinates. 
The calculation uses unwrapped atomic positions to account for periodic boundary conditions correctly.
\begin{equation}
    R_{g,s}^2 = \frac{1}{s}\sum_{s}^{*}|\vec{r_i}-\vec{R}_{\text{com}}|^2
\end{equation}
where $\vec{r_i}$ is the unwrapped coordinate of node $i$ and $s$ is the cluster size. This formula has an O(s) computational cost.

\item [d)] The correlation length $\xi$ is determined from the second moment of the gyration radius distribution $R_{g,s}$, weighted by cluster size, providing a measure of the characteristic cluster size.
\begin{equation}
    \xi^2 = \frac{\sum_s^* 2R_s^2 s^2 n_s(\phi)}{\sum_s^* s^2 n_s(\phi)}
\end{equation}

\item [e)] The percolation probability $\Pi(\phi)$ quantifies the likelihood that a cluster truly spans the simulation box through the periodic boundaries. 
A cluster percolates along a given direction when a breadth-first search (BFS) identifies a path that crosses the periodic boundaries of the simulation box and eventually loops back to its starting node.
This indicates the presence of a linearly independent period vector in that direction.
The detection is based on the rigorous period vector algorithm~\cite{livraghi_exact_2021}, which accurately identifies percolation by tracking the algebraic dimension of the cluster's periodic connections - distinguishing between one-dimensional, two-dimensional, and fully three-dimensional percolation. 
In Nexus-CAT, a cluster is considered percolating if it spans all three dimensions.

\item [f)]  The largest cluster size $S_\text{big}$ is the size of the single-largest cluster in the system for a given connectivity, regardless of whether it percolates.

\item [g)] In contrast to $S_\text{big}$, the spanning cluster size S$_\text{max}$ is the size of the largest finite (\textit{i.e.}, non-percolating) cluster. 
This is a crucial metric for analyzing the sub-critical regime and the fractal structure.

\item [h)]  Finally, the order parameter P$_\infty$ is the fraction of nodes in the percolating cluster. P$_\infty$ is a primary indicator in percolation and is equivalent to the order parameter in a standard phase transition.
\begin{equation}
    P_\infty = \begin{cases}0 & \text{if no cluster percolates} \\
    \frac{S_\text{big}}{N_{\text{clustered}}} & \text{if a cluster percolates}
\end{cases}
\end{equation}
where $N_{\text{clustered}}$ is the total number of nodes participating in clusters (of the same cluster connectivity).

\end{itemize}
  
\subsubsection{Outputs and Statistics}

Each \textit{Analyzer} block mentioned above stores the results of the calculated property of each frame, and it computes the average over the number of frames analyzed.
The program automatically prints the results of percolation properties along with their statistical uncertainties to files in a user-specified export directory in \textit{GeneralSettings}.
Additionally, users can export the unwrapped atomic coordinates of the identified clusters to XYZ files, facilitating direct visualization and subsequent analysis.
Once all results have been printed to their respective files, the program ends.

\section{Implementation details}
\label{sec:implementation_details}

\subsection{Launch script and settings details}
\label{sec:settings}
\sloppy
Now that we have an overview of the programme and how it works, this subsection will recapitulate and provide details about some additional settings as well as the contents of the launch script.
As mentioned earlier, this script uses a \textit{SettingsBuilder} to construct a comprehensive \textit{settings} object that consolidates all parameters required for an analysis.
The \textit{settings} object is composed of four distinct blocks, each managing specific aspects of the main function workflow: 
\begin{itemize}
    \item \textbf{GeneralSettings}: This block handles overall project configurations such as the project name, export directory, path to the trajectory file, range of frames to process, and options for verbose output, logging, and performance tracking. It also includes a flag to apply periodic boundary conditions.
    \item \textbf{LatticeSettings}: This section defines parameters related to the simulation box's lattice. It allows users to specify if a custom lattice should be applied or if the lattice information should be read directly from the trajectory file. We recall that the lattice format is a $3\times3$ matrix (with components $l_{xx}$, $l_{xy}$, $l_{xz}$, $l_{yx}$, ... ), which allows the code to handle general parallelepipeds.
    \item \textbf{ClusteringSettings}: This is a highly flexible block that dictates how atomic clusters are identified within any particle system. Users can define: \textit{node\_types}, a list of particle species (or type) of interest for the clustering analysis; and \textit{cutoffs}, inter-particle distance cutoffs needed for coordination number calculations. The latter uses the \textit{coordination\_mode} which can take different values including \textit{all\_types}, \textit{same\_type}, \textit{different\_type} or a specific \textit{node\_type}. The \textit{shared\_mode} works similarly. This comprehensive set of options, along with the previously mentioned clustering strategies, can be tailored to treat any atomistic system.
    \item \textbf{AnalysisSettings}: This block allows users to select which cluster properties are calculated. Options include the percolation properties, such as average cluster size, largest cluster size, concentration, spanning cluster size, gyration radius, correlation length, percolation probability, order parameter, and cluster size distribution. It also includes an \textit{overwrite} option to control whether existing analysis files are overwritten or appended, and an option to print out the unwrapped coordinates of the identified clusters. The latter has three \textit{print\_mode} options:  (i) \textit{all} all clusters in one file (frame separated),  (ii) \textit{connectivity} all clusters of the same connectivity in one file, and (iii) \textit{individual} each cluster has its own file. 
\end{itemize}

The modular design, complemented by the data structure and the builder pattern, provides a precise, robust, and validated method for configuring analyses in Nexus-CAT.

\subsection{Libraries dependency}
\label{sec:librairies}

The Nexus-CAT framework is implemented as a pure Python package designed for computational efficiency and ease of deployment.
The software architecture leverages several core scientific computing libraries to achieve optimal performance in percolation analysis of atomistic trajectories.
Note that each library and its appropriate version are installed alongside the Nexus-CAT installation.
\newline

\textit{NumPy} ($\geq$1.20.0)~\cite{Numpy} serves as the foundation for all numerical operations, providing efficient array manipulation and vectorized computations essential for handling large-scale trajectory data.
As already mentioned, the neighboring nodes are fetched using the function \textit{cKDTree} from the \textit{SciPy} package~\cite{virtanen_scipy_2020}.
Repetitive geometric functions for wrapping positions within the lattice, or for calculating distances or angles, are accelerated using \textit{Numba} ($\geq$0.53.0)~\cite{Numba}.
\textit{Numba} provides just-in-time (JIT) compilation of Python functions to machine code, achieving near-C performance for computationally intensive loops without compromising code readability or maintainability.
The toolkit incorporates \textit{psutil} ($\geq$5.8.0)~\cite{psutil} to monitor computer performance and store the data into a \textit{json} file.
User experience is enhanced through \textit{tqdm} ($\geq$4.50.0)~\cite{tqdm} for progress tracking during long-running analyses and \textit{colorama} ($\geq$0.4.4)~\cite{colorama} for cross-platform colored terminal output, providing clear visual feedback on analysis status and results.
The modular architecture supports Python versions $\geq$3.9, ensuring compatibility with modern Python environments.
The package adheres to standard Python packaging conventions using \textit{setuptools}, enabling straightforward installation via \textit{pip} and seamless integration into existing computational workflows. 
The minimal dependency footprint was deliberately chosen to reduce installation complexity while maintaining high performance, making the toolkit accessible to researchers across a range of computational environments, from personal workstations to high-performance computing clusters.
Installation instructions are available in the GitHub repository and the code documentation.

\subsection{Benchmarks and scalability}

To evaluate the program's scalability, we design a benchmark that can be analyzed with each clustering strategy.
We consider a three-dimensional simple cubic lattice of linear size $L$ with unit lattice spacing.
Two interpenetrating sublattices are defined: sublattice $\mathcal{A}$, with origin at $(0,0,0)$, hosts species A (network formers), while sublattice $\mathcal{B}$, shifted by $(0.5, 0.5, 0.5)$, hosts species B (bridging nodes).
Each sublattice contains $L^3$ sites, yielding a total of $2L^3$ available sites.
Sites on each sublattice are occupied randomly and independently with probability $\rho/2$, such that the overall occupation density is $\rho = N / (2L^3)$, where $N$ is the total number of occupied sites.
Species A and B are assigned randomly among the occupied sites.

Two densities are considered: $\rho = 0.4$, which lies below the percolation threshold and produces only finite clusters, and $\rho = 0.6$, which lies above the threshold and yields a percolating cluster coexisting with finite clusters.
For example, for $L = 50$, the total number of occupied nodes is $N = 50{,}000$ at $\rho = 0.4$ and $N = 75{,}000$ at $\rho = 0.6$.
Both configurations are generated for six system sizes: $L = 50, 60, 70, 80, 90, 100$.
\\ 

Each configuration of the system is analyzed with each clustering strategy as follows, and considering a single cutoff distance $\mathcal{A}-\mathcal{B}$ of $0.9$:
\begin{itemize}
    \item \textbf{DistanceStrategy}: Clusters connecting nodes $\mathcal{A}$ and $\mathcal{B}$ ($\mathcal{A}-\mathcal{B}$).
    \item \textbf{BondStrategy}: Clusters connecting  nodes $\mathcal{A}$ sharing a common neighbor $\mathcal{B}$ ($\mathcal{A}-\mathcal{B}-\mathcal{A}$).
    \item \textbf{CoordinationStrategy}: Clusters connecting  nodes $\mathcal{A}$ with the same coordination number $z$, with $z$ the number of $\mathcal{B}$ around a node $\mathcal{A}$ ($\mathcal{A}\mathcal{B}_z-\mathcal{A}\mathcal{B}_z$) and $z$ ranging from 1 to 6.
    \item \textbf{SharedStrategy}: Similarly to \textit{CoordinationStrategy} but adding a threshold number of shared $\mathcal{B}$ between nodes $\mathcal{A}$ ($\mathcal{A}\mathcal{B}_z*-\mathcal{A}\mathcal{B}_z$). 
\end{itemize}

As previously mentioned, Figure~\ref{fig:strategies} illustrates the four strategies applied to the \textit{v}-SiO$_2$ systems. The same strategies are applied to the benchmark lattice with generic species $\mathcal{A}$ and $\mathcal{B}$ (replacing Si and O, respectively) on the interpenetrating simple cubic sublattices. 

\begin{figure}[htbp]
    \centering
    \includegraphics[width=0.7\linewidth]{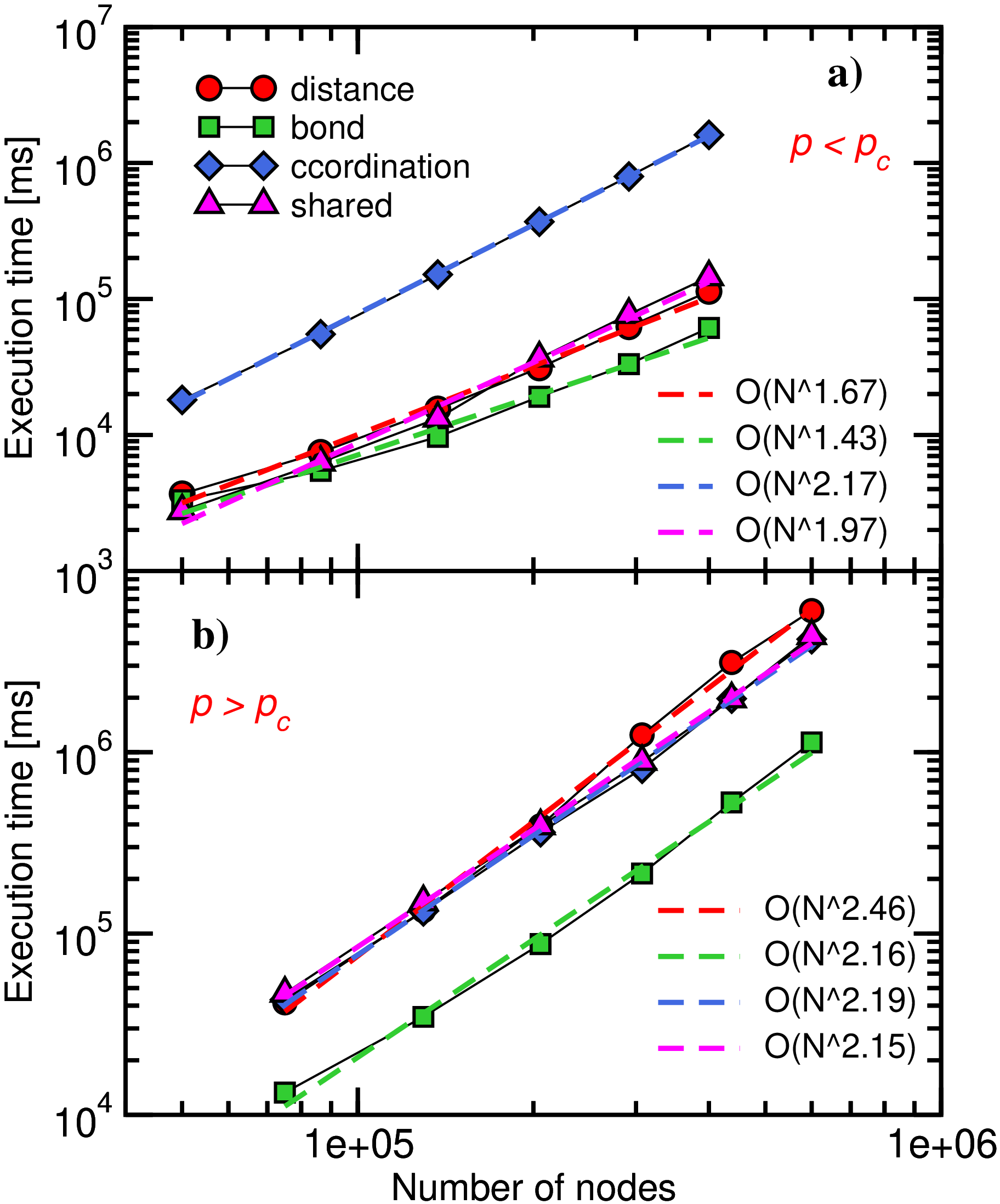}
    \caption{
    Performance benchmark of the Nexus-CAT package  \textit{DistanceStrategy} (red), \textit{BondStrategy} (green), \textit{CoordinationStrategy} (blue), and \textit{SharedStrategy} (magenta) methods. The red and blue symbols are the data, and the lines are the power-law fits.}
    \label{fig:benchmark}
\end{figure}

Figure~\ref{fig:benchmark} displays the execution time as a function of the number of sites $N$. 
This reveals the program's algorithm complexity to be approximately $\simeq O(N^{1.8})$ without a percolating cluster and $\simeq O(N^{2.2})$ with a percolating cluster within the range of 50k and 600k nodes. 
The complexity is expected to increase as the percolation order parameter approaches $1$ with increasing density. 

\subsection{Validation}
\label{sec:validation}
To validate the correctness of our implementation, we tested the package against well-established theoretical results for 3D site percolation on simple cubic lattices~\cite{stauffer_scaling_1979,xu_simultaneous_2014,borinsky_five-loop_2021,lorenz_precise_1998}.
Simulations were conducted on lattices of linear sizes $L = 20, 25, 30, 35, 40, 45$, and $50$ (i.e., $N = L^3$ sites), using the \textit{DistanceStrategy} with periodic boundary conditions.
For each system size, we generated a series of configurations with site occupation probability $p$ varying from 0.2 to 0.4 in increments of $\Delta p = 0.002$.
This protocol, which consists of scanning a control parameter across a transition, is deliberately chosen to mirror the typical workflow in atomistic simulations of disordered materials, where percolation transitions are driven by a continuously varying thermodynamic quantity such as pressure, temperature, or density~\cite{hasmy_percolation_2021,hasmy_unravelling_2024}.
For each value of $p$, 100 independent configurations were generated, yielding a total of 10{,}000 frames per lattice size.
Note that Nexus-CAT reads trajectories in the extended XYZ format, so each set of configurations was written as a multi-frame XYZ file; as a consequence, the statistical sampling remains modest compared with dedicated lattice percolation studies~\cite{xu_simultaneous_2014}.

\begin{figure}[H]
    \centering
    \includegraphics[width=\linewidth]{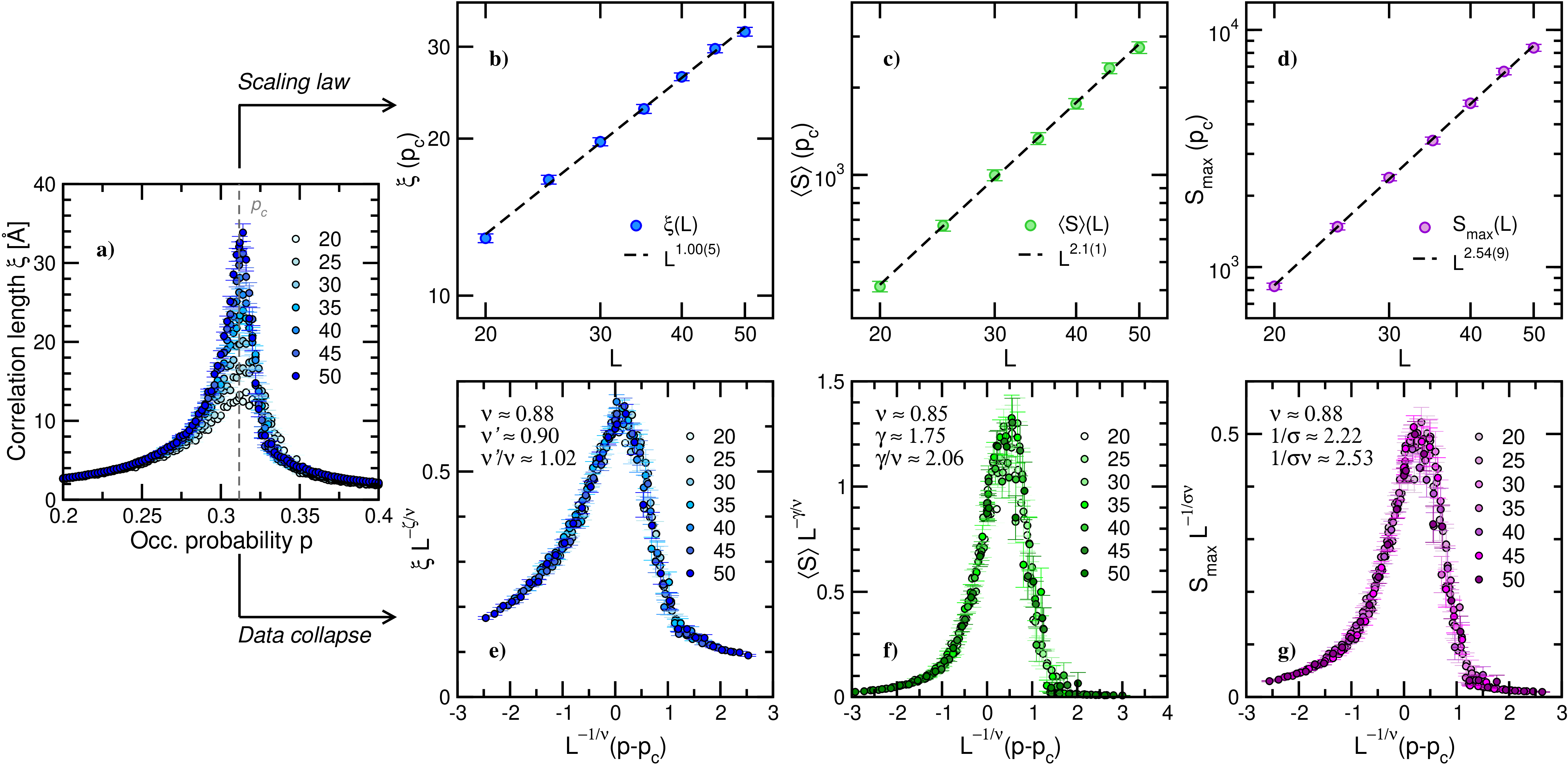}
    \caption{
        Finite-size scaling analysis of cluster properties. Scaling laws of \textbf{a)} the correlation length $\xi\propto L$, \textbf{b)} the average cluster size $\langle S\rangle\propto L^{\gamma/\nu}$, and \textbf{c)} the spanning cluster size S$_\text{max}\propto L^{D_f}$ where colored circles are the data at $p_c$ and the black dashed lines are the fits. Data collapse of \textbf{d)} the correlation length, \textbf{e)} the average cluster size, and \textbf{f)} the spanning cluster size. Note that the error on the critical exponents extracted via data collapse could not be estimated precisely.  
    }
    \label{fig:validation}
\end{figure}

Figure~\ref{fig:validation}.\textbf{a} shows the correlation length as a function of occupation probability $p$ for each system size; each curve depicts a maximum whose location tends to the theoretical percolation threshold for 3D site percolation on a simple cubic lattice $p_c \simeq0.3116$~\cite{stauffer_scaling_1979}. 
Via the finite-size scaling ansatz~\cite{binder_monte_2010}, one can extract the critical exponents at the percolation threshold if the condition $\xi\propto L$ is respected and compare with those reported in literature for the standard percolation universality class in $d=3$.  
Figure~\ref{fig:validation}.\textbf{b} displays the scaling law $\xi\propto L$ of the data shown in Fig.~\ref{fig:validation}.\textbf{a} which allow the ansatz and hence the calculation of the scaling laws in Figs.~\ref{fig:validation}.\textbf{c, d}: $\langle S\rangle\propto L^{\gamma/\nu}$, and S$_\text{max}\propto L^{D_f}$, respectively.

One can also extract the critical exponents individually by collapsing the data of each system size onto a single master curve via the function:
\begin{equation}
    \mathcal{A}(p,L) = L^{\zeta/\nu} \mathcal{F}(L^{-1/\nu}(p-p_c))
    \label{eq:data_collapse}
\end{equation}
with $\mathcal{A}$ the scaling quantity (\textit{i.e.}, $\langle S\rangle$, S$_\text{max}$ ...), and $\zeta$ the critical exponent (\textit{i.e.}, $\gamma$, $1/\sigma$ ...) associated with quantity $\mathcal{A}$.
Figures~\ref{fig:validation}.\textbf{e,f,g} show the data collapse of $\xi$, $\langle S\rangle$, S$_\text{max}$, respectively, allowing the extraction of the critical exponents which compare nicely with theoretical ones in Tab.~1 and thus validate the percolation properties calculation in Nexus-CAT.

\begin{table}[H]
\centering
\begin{tabular}{c|c|c}
&
\textbf{\textit{Nexus-CAT}}&
\textbf{Reference}\\
\hline
$\xi\propto L$ & 1.00(1) & 1.00 (th.~\cite{stauffer_scaling_1979}) \\
$\langle S\rangle\propto L^{\gamma/\nu}$ & 2.1(1) & 2.02 (calc.~\cite{borinsky_five-loop_2021}) \\
$S_\text{max}\propto L^{D_f}$ & 2.54(9) & 2.52~\cite{lorenz_precise_1998}\\
$\nu$ & 0.87 & 0.88~\cite{borinsky_five-loop_2021} \\
$\gamma$ & 1.75 & 1.78~\cite{borinsky_five-loop_2021} \\
$1/\sigma$ & 2.22 & 2.21~\cite{borinsky_five-loop_2021}\\
$\gamma/\nu$ & 2.06 & 2.02 (calc.~\cite{borinsky_five-loop_2021})\\
$1/\sigma\nu$ = $D_f$ & 2.53 & 2.52~\cite{lorenz_precise_1998}
\end{tabular}
\label{tab:exponents}
\caption{Critical exponents of standard site percolation ($d=3$) compared with those from Refs.~\cite{stauffer_scaling_1979,borinsky_five-loop_2021,lorenz_precise_1998}.}
\end{table}

\section{Case studies}

To demonstrate the versatility of Nexus-CAT across different classes of disordered materials, we present two applications corresponding to distinct clustering strategies.
The first concerns \textit{bonded} glasses, in which atoms are connected through covalent bonds: vitreous silica (\textit{v}-SiO$_2$), where silicon atoms are linked via bridging oxygens, and amorphous silicon (\textit{a}-Si), where silicon atoms are directly covalently bonded.
The second concerns a \textit{non-bonded} glass, amorphous ice (a-H$_2$O), analyzed within the framework introduced by O'Keeffe and Hyde~\cite{okeeffe_role_1981}, who showed that the distance between non-bonded first neighbor cations in many non-molecular crystals is nearly independent of the bridging anion, allowing the structure to be described solely through cation-cation (or, equivalently, anion-anion) arrangements.

In amorphous ice, each oxygen forms four hydrogen bonds with
neighbouring oxygens in the first coordination shell in
accordance with the Bernal–Fowler ice rules, and this
tetrahedral coordination is preserved across a wide range
of pressures. As pressure increases, however, some
oxygens from the second coordination shell are forced
into interstitial positions between the first and second
shells leading to an increase of the coordination number.
In that case, 
structural transformations are most naturally assessed using non-bonded criteria, with the oxygen-oxygen coordination number serving as the relevant structural descriptor.
These examples illustrate how the toolkit's modular strategy framework adapts to fundamentally different network topologies and bonding environments.

\subsection{Bonded glasses: vitreous silica and amorphous silicon}
\label{sec:bonded}

\subsubsection{Vitreous silica}
Vitreous silica forms a continuous random network of corner-sharing SiO$_4$ tetrahedra at ambient conditions.
Under compression, the network undergoes polyamorphic transitions characterized by the formation of higher-coordinated silicon species (SiO$_5$, SiO$_6$)~\cite{perradin_polyamorphism_2025,hasmy_percolation_2021}.

We employed the \textit{CoordinationStrategy} and the \textit{SharedStrategy} to analyze connectivity patterns of silicon-centered polyhedra.
Two silicon atoms are considered connected if they share an oxygen bridge. 
The coordination number of each silicon atom is determined by counting oxygen neighbors within the first coordination shell (cutoff distance of 2.3 Å for all pressures).

Our analysis focuses on four distinct percolation networks:
\begin{itemize}
    \item SiO$_4$-SiO$_4$ connectivity represents the tetrahedral network.
    \item SiO$_5$-SiO$_5$ connectivity characterizes intermediate-pressure structural changes.
    \item SiO$_6$-SiO$_6$ connectivity corresponds to high-pressure octahedral coordination.
    \item SiO$_6$*-SiO$_6$ connectivity corresponds to a very-high-pressure octahedral coordination units that share at least two oxygen neighbors, akin to the stishovite polymorph of crystalline silica \cite{bykova_metastable_2018}.
\end{itemize} 
  
\begin{figure}[H]
    \centering
    \includegraphics[width=0.7\linewidth]{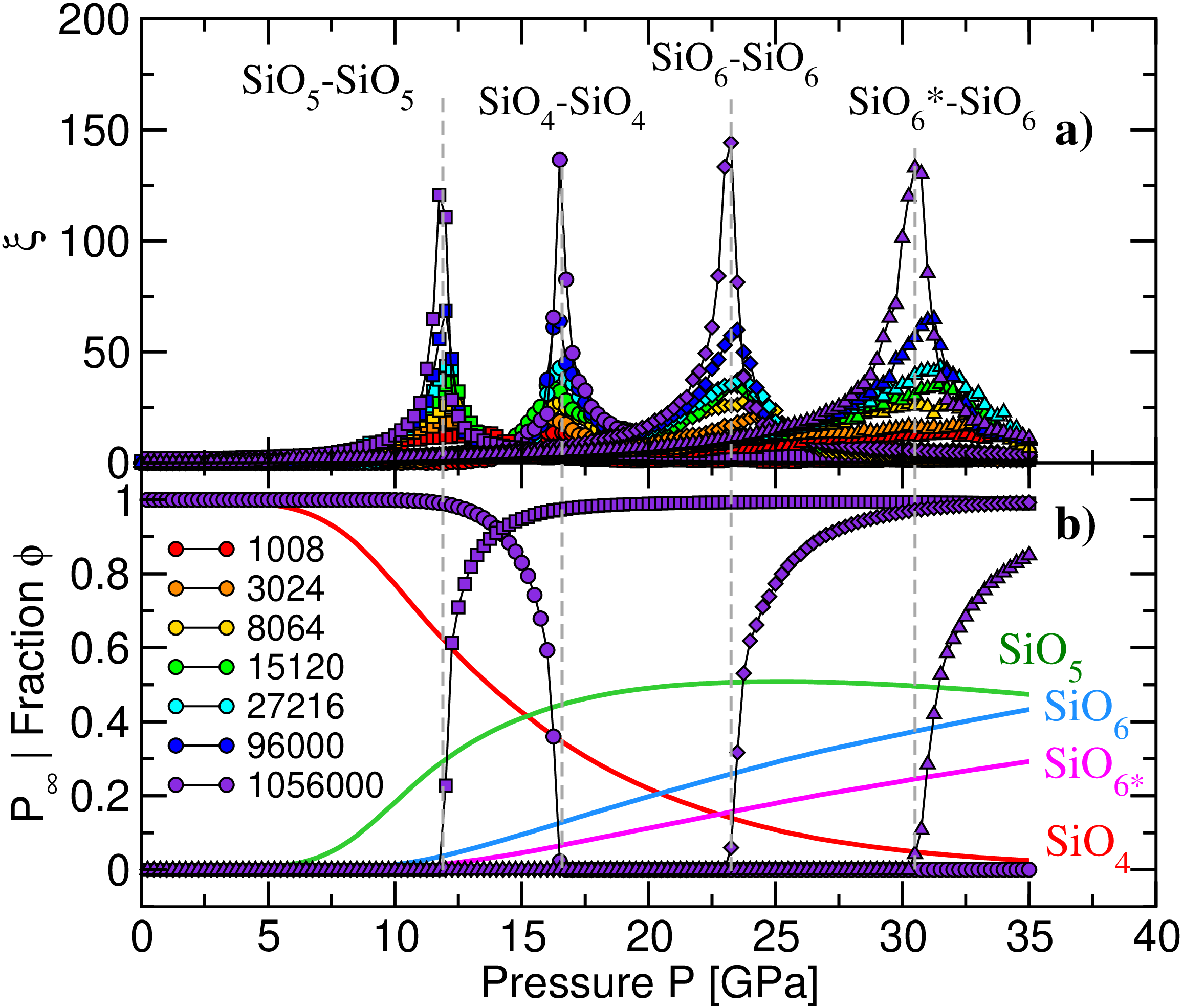}
    \caption{ Percolation transitions in \textit{v}-SiO$_2$ under compression. \textbf{a)} Correlation length $\xi$ of SiO$_4$-SiO$_4$ (circles), SiO$_5$-SiO$_5$ (squares), SiO$_6$-SiO$_6$ (diamonds), and SiO$_6$*-SiO$_6$ (triangles) with 7 system sizes $N=1008,3024,8064,15120,27216,96000$ and $1056000$ atoms in red, orange, yellow, lime, cyan, blue, purple, respectively. \textbf{b)} Order parameter P$_\infty$ (symbols) and SiO$_Z$ fractions $\phi$ (colored lines) as a function of pressure, $\phi^{(SiO_4)}$ in red, $\phi^{(SiO_5)}$ in lime, $\phi^{(SiO_6)}$ in blue, and $\phi^{(SiO_6*)}$ in magenta. Grey dashed lines are the percolation threshold of SiO$_Z$-SiO$_Z$, $P_c\simeq11.7$ GPa for $Z=5$, $P_c\simeq16.9$ GPa for $Z=4$, $P_c\simeq23.0$ GPa for $Z=6$ and $P_c\simeq30.0$ GPa for $Z=6*$. }
    \label{fig:vitreous_silica}
\end{figure}

Figure~\ref{fig:vitreous_silica} illustrates the evolution of percolation properties with pressure. 
The SiO$_4$ network exhibits a percolation threshold at $\phi_c^{(SiO_4)} \simeq 0.39$, corresponding to the onset of tetrahedral network breakdown. 
The emergence of SiO$_5$ and SiO$_6$ networks occurs at higher pressures, with thresholds $\phi^{(SiO_5)} \simeq 0.30$ and $\phi^{(SiO_6)} \simeq 0.27$, respectively.
Finally, the SiO$_6$ connected only via edge-sharing (\textit{i.e.} two oxygens) occurs at $\phi^{(SiO_6*)} \simeq 0.24$.
A detailed analysis was carried out in ~\cite{perradin_criticality_2025} to extract the critical exponents, following an approach similar to that described in Sec.~\ref{sec:validation}. The universality class of the transitions was then compared with that of standard percolation.

\subsection{Amorphous silicon}
\label{sec:non-bonded_aSi}

Amorphous silicon (\textit{a}-Si) is a prototypical single-species covalent network, predominantly composed of tetrahedrally coordinated Si atoms ($Z=4$) at ambient conditions.
Under compression, \textit{a}-Si undergoes structural transitions involving an increase in the average coordination number, analogous to the transformations observed in vitreous silica~\cite{deringer_origins_2021}.
Unlike \textit{v}-SiO$_2$, however, there is no bridging species: the Si-Si covalent bonds directly define the network topology.

Since \textit{a}-Si contains only one atomic species, the cluster analysis is performed using the \textit{CoordinationStrategy} (or equivalently, the \textit{DistanceStrategy}) applied directly to Si-Si pairs, with a cutoff distance of 3.0~\AA{}.
The local structural descriptor is the SiSi$_Z$ coordination number, \textit{i.e.}, the number of silicon nearest neighbors around each silicon atom.
Following the convention introduced for the non-bonded analysis of a-H$_2$O ~\cite{hasmy_unravelling_2024,perradin_criticality_2025}, we adopt the same classification of local amorphous structures: low-density (LDA) for $Z=4$, high-density (HDA) for $Z\in[5,7]$, and very-high-density (VHDA) for $Z\geq8$.

\begin{figure}[H]
    \centering
    \includegraphics[width=0.7\linewidth]{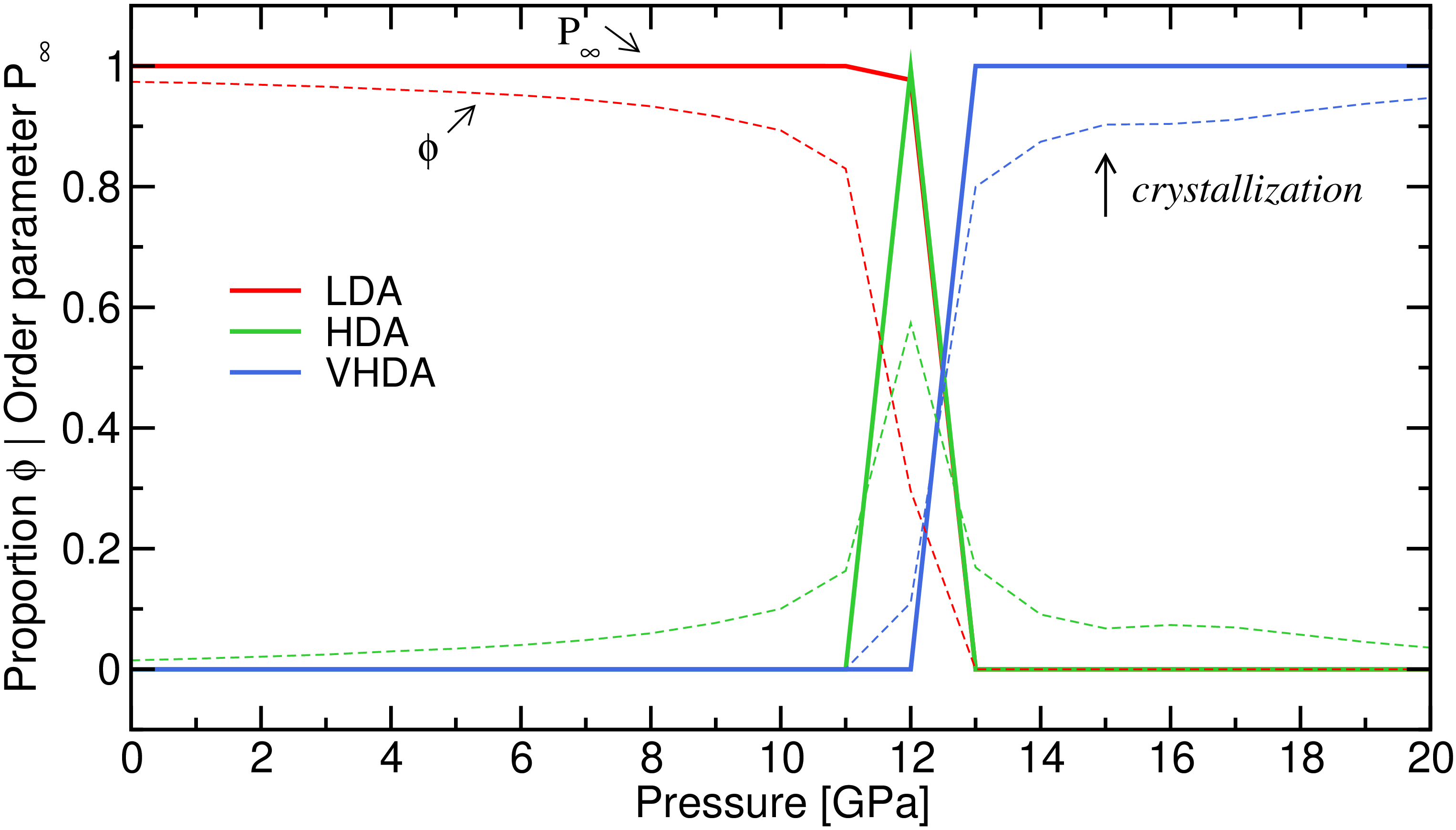}
    \caption{Percolation transitions in amorphous silicon under compression using original data from Deringer et al.~\cite{deringer_origins_2021}. Order parameter $P_\infty$ as a function of pressure for LDA ($Z=4$, red), HDA ($Z > 4$, green), and VHDA ($Z \geq 8$, blue) SiSi$_Z$ clusters. The dashed lines are the proportions of $Z=4$ coordinated Si in red , $Z > 4$ in green and $Z \geq 8$ in blue. The sharp sequence of transitions between 11 and 13~GPa coincides with the structural collapse reported in Ref.~\cite{deringer_origins_2021}.}
    \label{fig:amorphous_si}
\end{figure}

Figure~\ref{fig:amorphous_si} shows the pressure evolution of the order parameter $P_\infty$ for the three local amorphous structures with original data from Deringer et al.~\cite{deringer_origins_2021}.
These results provide the first observation of percolation transitions in a semiconducting glass.
At ambient pressure, the LDA network is fully percolated ($P_\infty \simeq 1$), reflecting the predominantly fourfold-coordinated covalent network of \textit{a}-Si.
Upon compression, the LDA percolating cluster persists up to approximately 11~GPa.
Between 11 and 13~GPa, a remarkably sharp sequence of percolation transitions occurs: the HDA cluster percolates, and almost simultaneously the LDA network depercolates, and the VHDA cluster emerges as the new system-spanning structure.
This narrow pressure window contrasts with the broader, more gradual transitions observed in v-SiO$_2$ and a-H$_2$O~\cite{perradin_criticality_2025}, and coincides precisely with the structural collapse reported by Deringer~\textit{et~al.}~\cite{deringer_origins_2021}, where a sudden volume reduction from $\sim$18 to $\sim$14~\AA$^3$/atom was observed at 12--13~GPa in GAP-driven molecular dynamics simulations of 100,000 atoms.
 In that study, the collapse leads to a transient VHDA phase, which subsequently crystallizes into a polycrystalline simple hexagonal structure above $\sim$15~GPa~\cite{deringer_origins_2021}.
It shows that the percolation transition brings the structure of $a$-Si to a metastable amorphous state sufficiently close to the potential energy minimum of the crystalline phase to enable an amorphous-to-crystal transition. Indeed, the Si-coordination of the VHD phase ($Z\geq 8$) is similar to that of the crystalline phase $Z = 8$~\cite{deringer_origins_2021}.
This result supports the assumption of Ref.~\cite{hasmy_percolation_2021}  that during a percolation transition, the structure of the percolating cluster is akin to that of a crystalline counterpart. In other words, both amorphous and crystalline states of $a$-Si around 11-13 GPa belong to the same megabasin of potential energy~\cite{machon_pressure-induced_2014}.
The sharpness of the transitions in \textit{a}-Si, as captured by Nexus-CAT, provides a complementary long-range structural perspective on this abrupt transformation, and illustrates the toolkit's ability to detect rapid polyamorphic changes in single-species covalent networks.

\subsection{Non-bonded glass: amorphous ice}
\label{sec:non-bonded_aH2O}

In amorphous ice (a-H$_2$O), the number of hydrogen bonds per molecule remains essentially unchanged with pressure, consistently maintaining fourfold coordination in accordance with the ice rules~\cite{hasmy_unravelling_2024}.
Consequently, structural transformations cannot be captured by tracking bond formation or breaking, and are instead assessed using non-bonded criteria based on the oxygen--oxygen coordination number $Z_{\mathrm{OO}}$.

\begin{figure}[H]
    \centering
    \includegraphics[width=0.7\linewidth]{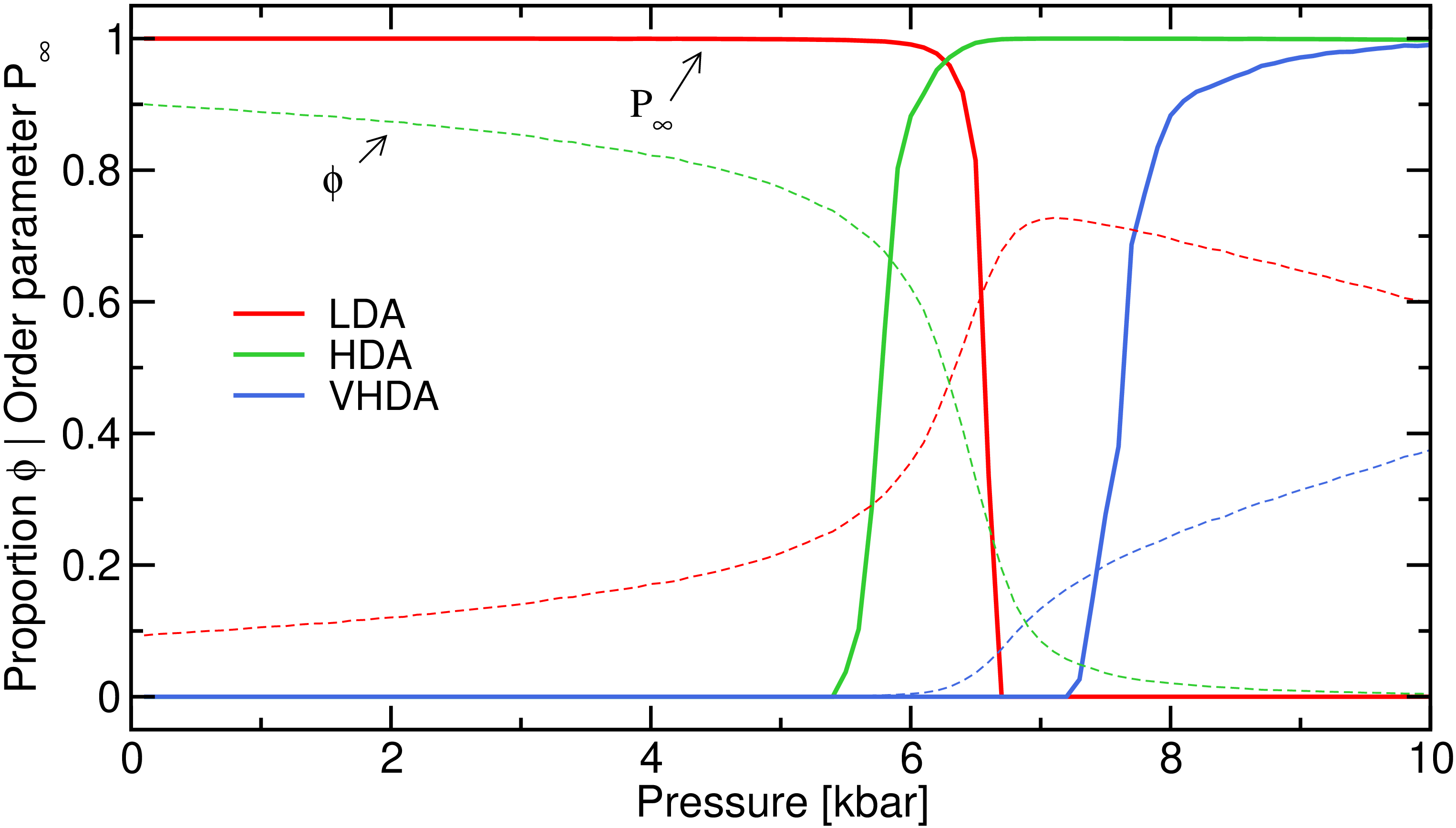}
    \caption{Percolation transitions in amorphous ice under compression at 124K. Full lines are the order parameters $P_\infty$ as functions of pressure for LDA ($Z=4$, red), HDA ($Z > 4$, green), and VHDA ($Z \geq 8$, blue) OO$_Z$ clusters, and dashed lines are the proportions of $Z=4$ coordinated O in red, $Z > 4$ in green and $Z \geq 8$ in blue.}
    \label{fig:ice}
\end{figure}

The cluster analysis is performed using the \textit{CoordinationStrategy} applied to oxygen--oxygen pairs, with a cutoff distance of 3.5~\AA{}.
The local structures are classified as follows: low-density (LDA) for $Z_{\mathrm{OO}}=4$, corresponding to nearly perfect tetrahedral arrangements characteristic of low-density amorphous ice; high-density (HDA) for $Z_{\mathrm{OO}}=5$--$7$, corresponding to distorted configurations; and very-high-density (VHDA) for $Z_{\mathrm{OO}} \geq 8$.
The cluster analysis does not distinguish between the various possible connectivities (\textit{e.g.}, corner-sharing, edge-sharing) of OO$_Z$ units. 
Figure~\ref{fig:ice} shows the pressure evolution of the order parameter $P_\infty$ for the three local amorphous structures in a-H$_2$O at 124~K. 
As in $a$-Si, we observe a sequence of percolation transitions: the LDA network progressively depercolates under compression, while the HDA and, subsequently, the VHDA clusters emerge and percolate. 
The transitions are more gradual than in \textit{a}-Si, consistent with the continuous nature of the structural rearrangements in amorphous ice under pressure.

This phenomenological model served as a unifying framework to describe metastable transformations in water across a broad pressure–temperature domain~\cite{hasmy_unravelling_2024}. The emergence, coexistence, and disappearance of LDA, HDA, and VHDA percolating clusters were tentatively associated with amorphous thermal phase transformations in the pressure-temperature diagram.

\section{Conclusion}
\label{sec:conclusion}

We have presented Nexus-CAT, an open-source Python package for cluster detection and percolation analysis of atomistic simulation trajectories.
The toolkit combines multiple physically motivated clustering strategies with the rigorous computation of percolation properties, and has been validated against established theoretical results for three-dimensional site percolation on simple cubic lattices. 
Its application in pressurized vitreous silica, amorphous Si, and amorphous ice demonstrates its ability to characterize successive percolation transitions of coordination-specific polyhedral networks across a wide range of system sizes.
In $v$-SiO$_2$, the critical exponents associated with the pressure-induced transitions from low- to highly coordinated percolating clusters are consistent with those of standard percolation~\cite{perradin_criticality_2025}. In $a$-H$_2$O, a parallel was made between the LD, HD, and VHD amorphous phases, and thermodynamic anomalies, {\it i.e.} the compressibility and the density~\cite{hasmy_unravelling_2024}. Finally, in $a$-Si, the results discussed here provide the first evidence of a percolation transition acting as a precursor to an amorphous-to-crystalline phase transition. 
These studies show that the ability of Nexus-Cat to define long-range descriptors and to track their growth up to percolation under various thermodynamic conditions opens new avenues for understanding the physical properties of disordered materials. 

More broadly, Nexus-CAT provides a general framework applicable to disordered systems, including multi-component glasses, gels, and cement, under varying external conditions, \textit{e.g.} temperature, pressure, and atomic or molar fractions.




Future developments of Nexus-CAT can be envisioned to significantly extend its physical relevance. More advanced geometric descriptors would improve the characterization of cluster morphology, and coupling the analysis to local energy or charge distributions would help relate long-range structure to thermodynamic and electronic properties in amorphous materials. Access to larger simulated systems ($>10^6$ elementary units, {\it e.g.} via  frame-level parallelism and GPU offloading of the Numba-compiled geometry kernels) would allow percolation phenomena to be studied at more realistic length scales, while time-resolved cluster tracking would connect structural organization to dynamical properties. 


\section*{Credit authorship contribution statement}

\textbf{Julien Perradin:} Writing – review \& editing, Writing – original draft, Visualization, Validation, Software, Methodology, Investigation, Formal analysis, Conceptualization; \textbf{Simona Ispas:} Writing – review \& editing, Conceptualization, Validation, Supervision; \textbf{Anwar Hasmy:} Writing – review \& editing, Conceptualization, Validation, Supervision; \textbf{Bernard Hehlen:} Writing – review \& editing, Conceptualization, Validation, Supervision.

\section*{Data availability}

The code is available in a public GitHub repository at \url{https://github.com/jperradin/nexus}, and the case study data of \textit{a}-Si are available from Ref.~\cite{deringer_research_2021}. For \textit{v}-SiO$_2$ and \textit{a}-H$_2$O, the data are available upon reasonable request.

\section*{Declaration of interests}

The authors declare that they have no known competing financial interests or personal relationships that could have appeared to influence the work reported in this paper.

\section*{Acknowledgments}

J.P. acknowledges financial support from the French Ministry of Higher Education and Research (MESR) through a doctoral fellowship.
J.P. would also like to thank Prof. Ricardo Paredes (Universidad Iberoamericana, Mexico City) for fruitful discussions.
This work was granted access to the HPC resources of IDRIS and CINES under allocations 2023-[A0140910788], 2024-[A0160910788], and 2025-[AD010910788R1], made by GENCI.





\bibliographystyle{elsarticle-num}
\bibliography{main.bib}







\end{document}